\documentclass[prd,twocolumn,aps]{revtex4}

\usepackage{url}
\usepackage{dsfont}
\usepackage{epsfig} 
\usepackage{amsmath} 
\usepackage{amsfonts}
\usepackage{amssymb}
\usepackage{color} 
\usepackage{graphicx}

\newcommand{\figsize}{\fontsize{6}{6pt}\selectfont} 
\newcommand{\bra}[1]{\langle #1|}
\newcommand{\ket}[1]{|#1\rangle}
\DeclareMathOperator{\tr}{tr}
\newcommand{\field}[1]{\mathbb{#1}}

\newcommand{\R}{\field{R}}

\newcommand{\pro}[1]{\ket{#1}\bra{#1}}
\newcommand{\id}{\mathds{1}}

\begin{document}

\title {Monitoring Quantum Oscillations with very small Disturbance}

\author{J\"urgen Audretsch
 Felix E.\ Klee and Thomas Konrad\\
Fachbereich Physik,
Universit\"at Konstanz, Fach M 674,\\
D-78457 Konstanz, Germany
PACS: 03.65.Ta, 03.67.-a,  42.50.-p}
\date{16.07.2004}

\begin{abstract}
  We present a new scheme to detect and visualize oscillations of a single
  quantum system in real time. The scheme is based upon a sequence of very weak
  generalized measurements, distinguished by their low disturbance and low
  information gain. Accumulating the information from the single measurements
  by means of an appropriate Bayesian Estimator, the actual oscillations can be
  monitored nevertheless with high accuracy and low disturbance. For this
  purpose only the minimum and the maximum expected oscillation frequency need
  to be known. The accumulation of information is based on a general derivation
  of the optimal estimator of the expectation value of a hermitian observable
  for a sequence of measurements.  At any time it takes into account all the
  preceding measurement results.
\end{abstract}

\maketitle

\section{Introduction}\label{sec:intro}

Consider a two-level quantum system where the probability to find the system in
a projection measurement on a specific level oscillates due to a periodically
time-dependent external potential. We propose a measurement scheme which allows
to monitor these oscillations by means of a sequence of consecutive
measurements carried out on a single two-level system. As in the case of the
detection of gravitational waves we assume that we are dealing with a one shot
experiment, i.e. the measurements have to be carried out on a single quantum
system and the experiment cannot be repeated in order to acquire more
measurement data. The proposed measurement scheme yields a real time record of
the actual oscillations, which are sometimes called Rabi oscillations. This is
done by appropriately estimating after each measurement in the sequence the
actual value of the oscillating probability.

In contrast to measurements in classical physics, quantum measurements have the
following prominent feature: The more information they provide, the more they
change the state of the measured system. The most precise measurements are von
Neumann projection measurements.  They project a quantum system in an
eigenstate of the measured observable, which represents in general a drastic
disturbance of the system's state.

There is a broader class of measurements called generalized measurements, which
can be realized by coupling the system in question to an ancilla system and
carrying out a projection measurement on the ancilla. Depending on the kind of
coupling and its strength these indirect measurements can exert an influence on
the system which ranges from very weak to very strong.  They can be elegantly
described in the POVM formalism \cite{BuschGrabowskiLahti95}.  In order to keep
the disturbance caused by a sequence of measurements low, we employ generalized
measurements with very weak influence. On the other hand, the weaker the
influence of these measurements is, the less information about the measured
system they convey. This disadvantage can be compensated by accumulating data
from all measurements. A corresponding data processing scheme is proposed
below.

The strength of the influence of a sequence of measurements can be determined
by considering the case where, apart from the measurements, no other dynamical
influence is present. In the example of the oscillating two-level system this
corresponds to screening or turning off the time-dependent potential. For an
appropriate sequence of measurements a measure for the strength of the
influence is then given by the ``decoherence time''.  That is the period after
which the coherences of the systems' state have decayed to $1/e$ of their
initial value (for qubits cp.\ \cite{AudretschDiosiKonrad02}). The undisturbed
dynamics of the system, on the other hand, can be characterized by the time
scale $T_R$, which is the period of the oscillations, if no measurements are
carried out.

Comparing the decoherence time $T_d$ to the period $T_R$, roughly three modes
or regimes of measurement can be distinguished \cite{AudretschKonradScherer01}
: (i) for $T_d\gg T_R$ the system evolves approximately according to its
undisturbed dynamics, i.e. the disturbing influence of the measurements is
comparatively small.  (ii) $T_d\approx T_R$, both dynamical influences are equally
strong. (iii) $T_d\ll T_R$, decoherence induced by the measurements dominates the
dynamics of the system.  Looking in mode (iii) at the systems dynamics in the
selective regime, i.e. given certain measurement outcomes, one finds quantum
jumps or in the limit of a continuous projection measurement the Quantum Zeno
effect, where the system freezes in an eigenstate of the measured observable.

In \cite{AudretschKonradScherer01} it was shown that a detection of the
oscillations of a two-level system under the influence of an external field
with reasonable accuracy and disturbance is possible employing mode (ii). A
physical realization of this measurement scheme was proposed by probing a
photon oscillating between two cavities by a sequence of Rydberg atoms
\cite{AudretschKonradScherer02}, based on an experiment of Haroche et al.
\cite{BruneHarocheRaimond92}.

In contrast to the latter investigations we want to show that the results can
be improved by working in mode (i). The advantage of mode (i) obviously is the
weak influence and thus the low disturbance inflicted by the measurements. On
the other hand mode (i) represents a challenge because there the measurement
results are only poorly correlated to the actual state evolution. In order to
overcome this difficulty we study optimal estimates in sections \ref{sec2} and
\ref{sec3}. The presented approach is rather general. We optimally estimate the
expectation value of an arbitrary hermitian observable of a quantum system with
finite dimensional Hilbert space, given the result of a generalized measurement
(which can also consist of a sequence of consecutive measurements). The result
can be applied to our special case to estimate the probability to find a
two-level system in a projection measurement on a specific level. This is done
in section \ref{sec:appl-track-an} after giving a brief description of this
scheme. The results of numerical simulations of our measurement scheme are
discussed. An appendix contains a recipe for these simulations with useful
formulae to abbreviate the computations.

\section{Estimator for mean values of observables}\label{sec2}
There probably exists considerably more literature on state-estimation than on
the estimation of mean values of physical quantities such as energy, position
or spin of quantum systems. Nevertheless there might be questions which do not
require the maximal knowledge of the statistics of all measurements that can be
carried out on a quantum system---as it is represented by the state of the
system---but rather the knowledge of a single physical property such as the
mean position of a quantum particle.  For example, when attempting to detect
gravitational waves, only the mean spacial distances between test masses have
to be estimated at different times \cite{Fritschel.et.al98}.  Of course the
calculation of the mean value of an observable as well as state determination
is not an issue if a large ensemble of identically prepared systems is
available to be measured. But for experiments with restricted resources and
especially in one-shot experiments estimation procedures become essential.

Let us consider the following task. Given a quantum system with $d$-dimensional
Hilbert space $\mathcal H$. After a POVM measurement with result $m$ the state
of the system reads
\begin{equation}
\label{evolution}
  \ket{\psi_m}= \frac{M_m\ket{\psi}}{\sqrt{p(m|\psi)}}\,,
\end{equation}
where $M_m$ is the Kraus operator corresponding to the measurement
result $m$, and 
\begin{equation}
p(m|\psi)= \bra{\psi}M^\dagger_m M_m\ket{\psi}
\end{equation}
represents the probability to obtain $m$, provided the state before the
measurement was $\ket{\psi}$. Let us assume that we know the result $m$ and the
respective Kraus operator $M_m$ but we do not know the initial state $\ket{\psi}$.
What is the best way to estimate the parameter
\begin{equation}
  \theta_m:= \bra{\psi_m}A\ket{\psi_m}\,,
\end{equation}
which represents the expectation value of the observable $A$ with respect to
the state $\ket{\psi_m}$?

It turns out to be rewarding to base the parameter estimation on the least
squared error criterion: The value $g_m$ is an optimal estimate of the
parameter $\theta_m$ if it minimizes the expected square of the error

\begin{equation}
E\left((\theta_m-g_m)^2\right) = \frac{\int (\theta_m-g_m)^2
  p(m|\psi)p(\psi)d\psi}{\int p(m|\psi)p(\psi)d\psi}\,.
\end{equation}
For the sake of simplicity we assume no prior knowledge about the initial state
$\ket{\psi}$, i.e. $p(\psi)=1$ and $d\psi$ is a normed measure over the set of all pure
states which is invariant under the action of the rotation group $SU(d)$.

Taking into account the linearity of the expectation value, it is easy to see
that in this case the optimal estimator $g_m$ is equal to the expected value of
$\theta_m$:
\begin{equation}\label{bla}
  \begin{split}
    E\left((\theta_m-g_m)^2\right)&= E\left((\theta_m)^2\right)-
    2gE(\theta_m)+g_m^2\\ 
    &= \left(E(\theta_m)-g_m\right)^2 +
    \mbox{Var}(\theta_m)\,,
  \end{split}
\end{equation}
where 
\begin{equation}
  \mbox{Var}(\theta_m)=E(\theta_m^2)-E(\theta_m)^2
\end{equation}
represents the variance of $\theta_m$. The right-hand side of (\ref{bla}) assumes
a minimum for $g_m=E(\theta_m)$. Such a value $g_m$ is also called the Bayesian
estimate (cp.\ \cite{Leonard01}).

Evaluating $g_m$ we obtain:
\begin{align}\label{estimate1}
    g_m &= \frac{\int \theta_m p(m|\psi)d\psi}{\int p(m|\psi)d\psi} \nonumber\\
    &= \frac{\int \bra{\psi}M^\dagger_m A M_m\ket{\psi} d\psi}{\int
      \bra{\psi} M^\dagger_m M_m\ket{\psi} d\psi}\nonumber\\
    &= \frac{\tr[M^\dagger_m A M_m]}{\tr[M^\dagger_m M_m]}\,.
\end{align}
This quantity represents the best estimate of the expectation value of
observable $A$ after one single generalized measurement with result $m$, if the
state before the measurement is completely unknown. It is the best estimate in
the sense that it leads to the least expected squared error.

Formula (\ref{estimate1}) can in particular be applied to estimate, after a
generalized measurement, the probability to find a system with two levels $0$
and $1$ on level $1$.  In this case $A$ should be chosen to be the projector on
level $1$ since the expectation value of this projector is equal to the desired
probability.

\section{Estimator for sequential measurements}\label{sec3}
In this section we derive an optimal estimator for a sequence of measurements.
For the sake of broad applicability we consider the general case of a sequence
of generalized measurements carried out on a single $d$-level system with
unknown Hamiltonian.

The single measurements with Kraus operators $N_{m_n}$ are carried out
consecutively on a single quantum system at times $t= n\tau$, where $n$ is an
integer. Here the number $m_n$ represents the result of the $n$-th measurement.
The single measurements are of duration $\delta \tau$ and during this time the motion
due to the system's Hamiltonian $H$ can be neglected (impulsive measurement
approximation).  Between two consecutive measurements, the system evolves
according to the unitary operator
\begin{equation}
U=\exp \frac{i}{\hbar} H\tau \,.
\end{equation}
For later convenience we express the unitary evolution by means of a unit
vector ${\mathbf k}\in {\R}^{d^2-1}$ with components $k_j$ and an angle $0\leq \phi<
2\pi$:
\begin{equation}
    U=\exp \frac{i}{\hbar} H\tau=\exp i{\mathbf k \cdot \mathbf e}\phi   \,,
\end{equation}
where ${\mathbf k \cdot \mathbf e}=\sum_j k_j e_j$, and the $e_j$ form
a complete set of generators of $SU(d)$.  The state
$\ket{\psi_{\mathbf m}}$ of the system after $n$ measurements with
results $(m_1,\ldots,m_n)\equiv {\mathbf m}$ is then given by equation
(\ref{evolution}) with Kraus operator
\begin{equation}
\label{Mm}
  M_{\mathbf m}= N_{m_n} U N_{m_{n-1}}U\ldots N_{m_1} U\,
\end{equation}
instead of $M_m$.

As above we assume that we know the results of the $n$ measurements and,
consequently, the corresponding Kraus operators $N_{m_1},\ldots, N_{m_n}$, but we
know neither the unitary evolution $U$ between the measurements nor the initial
state $\ket{\psi}$ of the system. Our ignorance about $U$ has to be incorporated
into the optimal estimate of the observable $A$ after the $n$-th measurement.
Instead of only averaging over all possible initial states as in
(\ref{estimate1}) we also have to average over all possible unit vectors
$\mathbf k$ and possible angles $\phi$ weighted by the corresponding probabilities
$p({\mathbf k})$ and $p(\phi)$: 
\begin{equation}
  g_{\mathbf m} =\frac{\int \theta_{\mathbf m} p({\mathbf m}|\psi,{\mathbf k}, \phi ) p({\mathbf k})
    p(\phi)d\psi d^{\tiny d^2-1}{\mathbf k}d\phi }{\int p({\mathbf m}|\psi,{\mathbf k}, \phi)
    p({\mathbf k}) p(\phi)d\psi d^{\tiny d^2-1}{\mathbf k}d\phi}\,.
\end{equation}
We assume that the direction of ${\mathbf k}$ and the angle $\phi$ are equally
distributed, i.e. $p({\mathbf k})$=const and $p(\phi)$= const. The optimal
estimator $g_{\mathbf m}$ is then given by
\begin{eqnarray}
\label{estimator2}
  g_{\mathbf m} &=& \frac{\int \theta_m p(m|\psi,{\mathbf k}, \phi )d\psi
    d^{\tiny d^2-1}{\mathbf k}d\phi }{\int p(m|\psi,{\mathbf k}, \phi  )d\psi
    d^{\tiny d^2-1}{\mathbf k}d\phi}\,, \nonumber\\
    &=& \frac{\int  \tr[M^\dagger_{\mathbf m} A M_{\mathbf m}]d^{\tiny d^2-1}{\mathbf k}d\phi
    }{\int \tr[M^\dagger_{\mathbf m} M_{\mathbf m}]d^{\tiny d^2-1}{\mathbf k}d\phi}\,.
\end{eqnarray}
For unknown unitary evolution, $g_{\mathbf m}$ represents the estimate of the
expectation value of observable $A$ after $n$ measurements with results
${\mathbf m}=(m_1,\ldots,m_n)$. It minimizes the expected squared error.
Calculating $g_{\mathbf m}$ after each measurement in a sequence of
measurements yields an optimally updated estimate of the current mean value of
observable $A$.
 
\section{Application: Tracking  an oscillating qubit\label{sec:appl-track-an}}
We will now apply the estimator given in equation (\ref{estimator2})
to the 
sequential measurement of an oscillating qubit. Oscillating qubits are realized
for example by two-level systems such as coupled quantum dots, trapped atoms in
an external field or photons oscillating between two microwave cavities
\cite{AudretschKonradScherer02}. For the sake of concreteness, we consider a
two-level atom under the influence of a resonant laser field. The Hamiltonian
of such an atom can be approximated by
\begin{eqnarray}
  \label{H}
  H &=& E_0 \pro{0} +E_1\pro{1} \\
  && + \frac{\hbar\Omega_R}{2}(\ket{1}\bra{0}\exp\{-i\omega t\}
  +\ket{0}\bra{1}\exp\{i\omega t\})\,, \nonumber  \nonumber  
\end{eqnarray}
where $\omega = (E_1-E_0)/\hbar$, and the Rabi frequency $\Omega_R$
represents the strength of the coupling between the atom and the
electromagnetic field. The resulting motion of a state that is not
subjected to measurement is represented by
\begin{equation}
  \label{psi}
  \ket{\psi(t)}=c_0(t)\ket{0}+c_1(t)\ket{1}
\end{equation}
with $|c_1|^2=\frac{1}{2}(1+a \cos(\Omega_R t+\varphi))$ where the constants $a$ and $\varphi$
depend on the initial state of the qubit.

In order to observe the actual behaviour of the system in real time, we have to
measure and estimate the expectation value of $\pro{1}$, which is equal to
$|c_1(t)|^2$. In the following we take the viewpoint that we already know the
Hamiltonian (\ref{H}) apart from the precise value of the coupling strength
$\Omega_R$. This is quite a natural assumption, since the form of $H$ has to be
known in order to design measurements, which requires the knowledge of how to
couple a meter to the system.

In a first step we choose the Kraus operators. To obtain information about the
observable $\pro{1}$, it seems natural to employ measurements the  effects
($N_{m}^\dagger N_{m}$) of which commute with $\pro{1}$ (cp. \cite{konrad03}).
Because such effects are diagonal with respect to the basis states $\ket{0}$
and $\ket{1}$, i.e.\ 
\begin{equation}
  N_{m}^\dagger N_{m}= p_0^{(m)}\pro{0} + p_1^{(m)}\pro{1}\,,
\end{equation}
the probability to obtain the result $m$ depends directly on $|c_1(t)|^2$:
\begin{eqnarray}
p(m |\psi(t))&= &\bra{\psi(t)}N^\dagger_{m} N_{m}\ket{\psi(t)}\,, \nonumber\\
             &=& p_0^{(m)}+ \Delta p^{(m)} |c_1|^2\,,
\end{eqnarray}
where $\Delta p^{(m)}:= p_1^{(m)}-p_0^{(m)}$.  For the sake of simplicity we
consider measurements with two possible results, $+$ and $-$, and Kraus
operators $N_\pm= U_\pm \sqrt{N^\dagger_{\pm} N_{\pm}}$ with trivial unitary part, i.e.\ 
$U_\pm=\id$. The Kraus operators of each single measurement thus read
\begin{eqnarray}
\label{defN+}
N_+ &:=& \sqrt{p_0^+}\pro{0}+\sqrt{p_1^+}\pro{1}\,, \\
\label{defN-}
N_- &:=& \sqrt{p_0^-}\pro{0}+\sqrt{p_1^-}\pro{1} \,
\end{eqnarray}
with positive numbers $p_j^\pm$ which satisfy $p_j^+ +p_j^- =1$.  A detailed
analysis of optimal Kraus operators from the viewpoint of Bayesian estimates
will be presented elsewhere.

The change of state caused by consecutive measurements with Kraus operators
$N_\pm$ as given in Eqs.\ (\ref{defN+}-\ref{defN-}) can be quantified by the
decoherence time $T_d$. That is the average period after which the off-diagonal
elements $\bra{i}\rho(t)\ket{j}$ with $i\not=j\in \{0, 1\}$ of the systems density
operator $\rho(t)$ are decayed to $1/e$ of their original value, if the only
dynamical influence is given by the measurements. $T_d$ is related to the
decoherence rate $\gamma$ \cite{AudretschDiosiKonrad02} by
\begin{equation}
  T_d = \frac{2}{\gamma}= \frac{8\tau {\bar p}(1-{\bar p})}{(\Delta p)^2}
\end{equation}
with $\bar p:=(p_0^++p_1^+)/2$ and $\Delta p := p_1^{(+)}-p_0^{(+)}$. $\tau$ is the
time which passes between two consecutive measurements.  If unitary dynamics
generated by the Hamiltonian $H$ (\ref{H}) are present, they can create
coherences representing a counter weight to the decoherence caused by the
measurements.  For
\begin{equation}\label{mode(i)}
T_d\gg T_R\,,
\end{equation}
where $T_R=2\pi/\Omega_R$ is the period of the Rabi oscillation, the influence of the single measurements on the state of the system
becomes negligibly small as compared to the influence of the unitary dynamics.
In this mode, which was called mode (i) in the introduction, we run our
sequential measurement.

Apart from condition (\ref{mode(i)}) there is another
requirement for the sequence of measurements: a reasonably high number
of measurements should take place on the time scale $T_R$ of the
unitary dynamics in order to resolve these dynamics. Hence,
\begin{equation}\label{cond2}
T_R\gg \tau \,.
\end{equation}
According to our experience based on numerical simulations $T_R/\tau \approx O(10)$ is
sufficient to obtain good results (see below). Note that in order to meet
conditions (\ref{mode(i)}) and (\ref{cond2}) only a vague knowledge about the
order of magnitude of the time scale $T_R$ and accordingly of $\Omega_R$ is
necessary. Knowing an upper bound $T_R^>$ and a lower bound $T_R^<$ for $T_R$,
both conditions can always be satisfied by first inserting the lower bound
$T_R^<$ into (\ref{cond2}) and choosing $\tau$ accordingly. Having fixed the value
of $\tau$, the parameters $p_0$ and $p_1$ can be tuned such that
\begin{equation}
8 {\bar p}(1-{\bar p})/(\Delta p)^2\gg T_R^>/\tau\,.
\end{equation}
In other words: the sequential measurement has to consist of frequent
measurements which are sufficiently weak.

Since the influence of the 
measurements obeying (\ref{mode(i)})  is very weak on the time scale  
$T_R$ they convey only very little
information about the system over the period $T_R$. This is where the accumulation of
information by means of the Bayesian
estimate (\ref{estimator2}) comes into play. The estimate after the
$n$-th measurement makes the best possible  use of the data collected in the
previous $n-1$ measurements according to the least square criterion. 

In our special case the Bayesian estimate given in (\ref{estimator2}) reduces
to
\begin{equation}\label{eq:1}
  g= \frac{\int  \tr[M^\dagger_{\mathbf m} A M_{\mathbf m}]d\phi
    }{\int \tr[M^\dagger_{\mathbf m} M_{\mathbf m}]d\phi}\,
\end{equation}
with Kraus operators $M_{\mathbf m}$ given by Eq.\ (\ref{Mm}),(\ref{defN+}) and
(\ref{defN-}). The unitary evolution $U$ in $M_{\mathbf m}$ (cp.\ Eq.\ 
(\ref{Mm})) is most easily represented in the interaction picture, where
Hamiltonian (\ref{H}) reads $H_I=\hbar\Omega_R\sigma_x/2$ with the Pauli spin operator
$\sigma_x:=\ket{1}\bra{0}+ \ket{0}\bra{1}$:
\begin{equation}
U=\exp(-i\frac{\sigma_x\phi}{2})\,.
\end{equation}
Here the angle $\phi$ of rotation on the Bloch sphere is given by 
\begin{equation}
\phi=\Omega_R\tau=\frac{2\pi\tau}{T_R}\,.
\end{equation}

Because of condition (\ref{cond2}), which guarantees the temporal resolution of
the Rabi oscillations by the sequence of measurements, the angle $\phi$ is in fact
very small. We did not include this a priori information into the estimate
$g_{\mathbf m}$ used in our numerical simulations. Instead, we let $\phi$ run from
$0$ to $2\pi$ in the integral in equation (\ref{estimator2}).  This corresponds
to a completely unknown Rabi period $T_R$. An appropriate change of the range
of integration might lead to an improvement of the estimate $g_{\mathbf m}$.

Note that the representation of the Kraus operators $N_\pm$ in the interaction
picture is the same as in the Schr\"odinger picture used so far:
\begin{equation}
 N_\pm^{(I)}(t)= e^{\frac{i}{\hbar}H_0 (t-t_0)}N_\pm  
 e^{-\frac{i}{\hbar}H_0 (t-t_0)}=N_\pm\,
\end{equation}
with $H_0:=E_0\pro{0}+E_1\pro{1}$.

\begin{figure}
  \begin{center}
    \begin{picture}(0,0)%
      \includegraphics{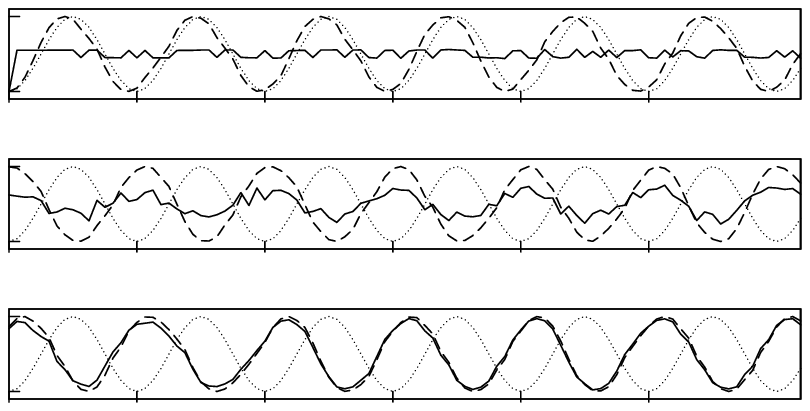}%
    \end{picture}%
    \begingroup
    \setlength{\unitlength}{0.0200bp}%
    \begin{picture}(10800,7559)(0,0)%
      \put(-150,5508){\makebox(0,0)[r]{\strut{}\figsize 1}}%
      \put(-150,4428){\makebox(0,0)[r]{\strut{}\figsize 0}}%
      \put(9212,4020){\makebox(0,0){\strut{}\figsize 5}}%
      \put(7370,4020){\makebox(0,0){\strut{}\figsize 4}}%
      \put(5527,4020){\makebox(0,0){\strut{}\figsize 3}}%
      \put(3685,4020){\makebox(0,0){\strut{}\figsize 2}}%
      \put(1842,4020){\makebox(0,0){\strut{}\figsize 1}}%
      \put(0,4020){\makebox(0,0){\strut{}\figsize 0}}%
      \put(11100,4020){\makebox(0,0){\strut{}\figsize $t/T_\text{R}$}}%
      \put(-150,3348){\makebox(0,0)[r]{\strut{}\figsize 1}}%
      \put(-150,2268){\makebox(0,0)[r]{\strut{}\figsize 0}}%
      \put(9212,1860){\makebox(0,0){\strut{}\figsize 77}}%
      \put(7370,1860){\makebox(0,0){\strut{}\figsize 76}}%
      \put(5527,1860){\makebox(0,0){\strut{}\figsize 75}}%
      \put(3685,1860){\makebox(0,0){\strut{}\figsize 74}}%
      \put(1842,1860){\makebox(0,0){\strut{}\figsize 73}}%
      \put(0,1860){\makebox(0,0){\strut{}\figsize 72}}%
      \put(11100,1860){\makebox(0,0){\strut{}\figsize $t/T_\text{R}$}}%
      \put(-150,1188){\makebox(0,0)[r]{\strut{}\figsize 1}}%
      \put(-150,108){\makebox(0,0)[r]{\strut{}\figsize 0}}%
      \put(9212,-300){\makebox(0,0){\strut{}\figsize 132}}%
      \put(7370,-300){\makebox(0,0){\strut{}\figsize 131}}%
      \put(5527,-300){\makebox(0,0){\strut{}\figsize 130}}%
      \put(3685,-300){\makebox(0,0){\strut{}\figsize 129}}%
      \put(1842,-300){\makebox(0,0){\strut{}\figsize 128}}%
      \put(0,-300){\makebox(0,0){\strut{}\figsize 127}}%
      \put(11100,-300){\makebox(0,0){\strut{}\figsize $t/T_\text{R}$}}%
    \end{picture}%
    \endgroup
  \end{center}
  \caption{Comparison between simulated evolution of a qubit's
  Rabi oscillations  and processed measurement signal for $\bar p=0.5$,
  $\Delta p=0.1$ and $\tau=T_R/16$.  Dashed curve:
  $|c_1|^2$ over time (in units of the Rabi period $T_R$) in the
  presence of weak measurements. Dotted curve: $|c_1|^2$
  over time in the absence of measurements. The solid curve corresponds to the evolution of the
  estimate $g$ based on the measurement results.\label{fig1}}
\end{figure}

\section{Results and Conclusion}

In Fig.\ref{fig1} the result from a numerical simulation of a sequential
measurement as specified above is plotted. The simulation started with the
initial state $\ket{\psi(t_0)}=\ket{1}$. The solid curve represents the evolution
of the estimate $g$ and the dashed curve corresponds to the dynamics of
$|c_1|^2$, taking into account the influence of the measurements. This
represents what really happens. The dotted curve displays how $|c_1|^2$ would
have evolved without measurements.

In the beginning of the sequential measurement(upper picture in Fig.\ref{fig1})
the curves of $|c_1|^2$ with and without measurements are close; thus the
disturbance due to the measurements is small. A small phase shift of the
oscillations however is recognizable. The values of the estimate $g$ are not
well correlated to the values of $|c_1|^2$. After many measurements (middle
picture in Fig.(\ref{fig1})) the estimate $g$ starts to approximate the
disturbed $|c_1|^2$ values, while the phase shift between the latter and the
values of $|c_1|^2$ without measurements has increased. The amplitude of the
oscillation of $|c_1|^2$ is not changed by the measurements if---as seen
here---it equals one in the absence of measurements. Eventually (lower picture
of Fig.\ref{fig1}) estimate $g$ and the curve of actual values of $|c_1|^2$
nearly coincide.

The main results of the simulations are: i) With increasing time the
estimate $g$ reflects the actual oscillations with growing
fidelity. After approximately hundred  Rabi cycles these   oscillations  
are monitored by $g$ with high accuracy. ii) In the presence of the weak measurements the sinusoidal shape and the period of the Rabi
oscillation is almost the same as in the absence of
measurements. The measurements however cause a phase shift of the
oscillation. Therefore the estimate $g$ also reflects shape and period of
the undisturbed Rabi oscillation. 

This demonstrates that our approach allows to monitor the periodic evolution of an expectation value with
high fidelity. The key to the monitoring are  measurements
with very low disturbance combined with an estimator which accumulates
at any time the information gained in the  sequence of  previous measurements.

\section{Appendix}

Simulations of the tracking procedure explained in section
\ref{sec:appl-track-an} were performed with a program \footnote{The program's
  source code is available from the authors upon request.} that is based on the
following algorithm.
\begin{enumerate}
\item Initialize the qubit's state vector $\ket{\psi}$ and the number of measurements,
  $n_{\text{max}}$. Set $n=1$.
\item Evolve $\ket{\psi}$ in time: $\ket{\psi}\to e^{-iH\tau/\hbar}\ket{\psi}$.\label{item:1}
\item Perform measurement:
  \begin{enumerate}
  \item Generate a (pseudo) random number $m_n$ whose value is either $0$ or $1$, 
    depending on the probability $p_{m_n}=\bra{\psi}N_{m_n}^\dagger N_{m_n}\ket{\psi}$.
  \item Update $\ket{\psi}$: $\ket{\psi}\to(N_{m_n}/\sqrt{p_{m_n}})\ket{\psi}$.
  \end{enumerate}
\item Calculate the estimator according to formula (\ref{eq:1}).
\item If $n<n_{\text{max}}$, then continue at step \ref{item:1} and increment $n$ by 1.
\end{enumerate}
In the program, the estimator is calculated using a variation of formula 
(\ref{eq:1}) that does not contain any integrals. Before specifying this 
variation, let us introduce the abbreviation $N_{jk}=\bra{j}N_{m_n}\ket{k}$ and the 
coefficient $\delta_{lm}^k$ which is $1$ for $l\leq k\leq m$ and $0$ otherwise. Then, the following
expression is equal to (\ref{eq:1}) for all Hamiltonians (\ref{H}) and 
\emph{all} sets of measurement operators (proof is omitted):

\begin{equation} 
  g=\frac{F_{11}}{F_{00}+F_{11}}\text{, }F_{jj}=\sum_{k=0}^{n-1}a_{jj(2k)}^{(n)}b_{k(n-1-k)}
\end{equation}
with
\begin{equation} 
  b_{kl}=\frac{2}{(k+l)!}\Gamma(k+\frac{1}{2})\Gamma(l+\frac{1}{2})
\end{equation}
and the recursive relation
\begin{equation} 
  \begin{split}
    a_{jlk}^{(n=0)}&=0\text{, }a_{jlk}^{(n=1)}=\sum_{p=0}^1N_{jp}N_{lp}^*\text{,}\\
    a_{jlk}^{(n\geq2)}&=\sum_{p=0}^1\sum_{q=0}^1\Bigl(\delta_{2(2n-2)}^kN_{jp}N_{lq}^*a_{pq(k-2)}^{(n-1)}\\
    +&\delta_{1(2n-3)}^ki(N_{j(1-p)}N_{lq}^*-N_{jp}N_{l(1-q)}^*)a_{pq(k-1)}^{(n-1)}\\
    +&\delta_{0(2n-4)}^kN_{j(1-p)}N_{l(1-q)}^*a_{pqk}^{(n-1)}\Bigr)\text{.}
  \end{split}
\end{equation}
Note, that in order to avoid convergence problems caused by low precision 
floating point data types, the program uses data types provided by the GMP 
library \footnote{The GNU Multiple Precision Arithmetic Library is available 
from \texttt{http://www.swox.com/gmp/}.}
for the calculation of the estimator.


\begin{thebibliography}{1}

\bibitem{BuschGrabowskiLahti95}
P.~Busch, M.~Grabowski, and J.P. Lahti.
\newblock {\em Operational Quantum Physics}.
\newblock Springer Verlag, Heidelberg, 1995.

\bibitem{AudretschDiosiKonrad02}
J.~Audretsch, L.~Di{\'o}si, and Th. Konrad.
\newblock Evolution of a qubit under the influence of a succession of weak
  measurements.
\newblock {\em Phys.\ Rev.}, A 66:022310(1--11), 2002.
\newblock E-print quant-ph/0201078.

\bibitem{AudretschKonradScherer01}
J.~Audretsch, Th. Konrad, and A.~Scherer.
\newblock A sequence of unsharp measurements enabling a real-time visualization
  of a quantum oscillation.
\newblock {\em Phys.\ Rev.}, A 63:052102, 2001.

\bibitem{AudretschKonradScherer02}
J.~Audretsch, Th. Konrad, and A.~Scherer.
\newblock Quantum optical weak measurements can visualize photon dynamics in
  real time.
\newblock {\em Phys.\ Rev. A}, 65:033814(1--6), 2002.

\bibitem{BruneHarocheRaimond92}
M.~Brune, S.~Haroche, J.M. Raimond, L.~Davidovich, and N.~Zagury.
\newblock Manipulation of photons in a cavity by dispersive atom-field
  coupling: Quantum-nondemolition measurements and generation of \lq\lq
  {S}chr\"odinger cat'' states.
\newblock {\em Phys.\ Rev.}, A 45:5193, 1992.

\bibitem{Fritschel.et.al98}
P.~Fritschel and et~al.
\newblock High power interferometric phase measurement limited by quantum noise
  and application to detection of gravitational waves.
\newblock {\em Phys.\ Rev.\ Lett.}, 80:3181--3184, 1998.

\bibitem{Leonard01}
T.~Leonard and J.S.J. Hsu.
\newblock {\em {B}ayesian methods: an analysis for statisticians and
  interdisciplinary researchers}.
\newblock Cambridge Univ.\ Press, Cambridge, U.K., 2001.

\bibitem{konrad03}
Th. Konrad.
\newblock {\em Less is more: on the Theory und Application of Unsharp
  Measurements in Quantum Mechanics}.
\newblock Phd-thesis, University of Konstanz, Germany, 2003.
\newblock URN:\ \url{urn:nbn:de:bsz:352-opus-10504}, URL:\
  \url{http://www.ub.uni-konstanz.de/kops/volltexte/2003/1050/}.

\end{thebibliography}

\end{document}